\documentclass[a4paper,11pt]{article}
\usepackage[latin1]{inputenc}
\usepackage[T1]{fontenc}
\usepackage{verbatim}
\usepackage{graphicx}
\usepackage{latexsym}
\usepackage{epsfig}
\usepackage{caption}

\makeatletter
\renewcommand{\fnum@figure}{\footnotesize\textbf{\figurename~\thefigure}}
\makeatother

\title{Detecting overlapping communities in linear time with P\&A algorithm}
\author{
Nicolas Pissard and Houssem Assadi
\\ 
\textit{\footnotesize{nicolas.pissard@francetelecom.com - houssem.assadi@francetelecom.com}}
\\
\\
\textit{\footnotesize{France Telecom R\&D, Issy-les-Moulineaux, France}}
}
\date{}

\begin{document}

\maketitle

\abstract{
This paper describes a new algorithm - P\&A algorithm - utilized in identifying overlapping communities in non oriented valued graph regardless of their number or their size. The complexity of this algorithm is minimal in the matter that the number of operations grows linearly with the number of vertices.

\section{Introduction}
Several algorithms have already been proposed in order to calculate graph partitions. These algorithms can be classified in two categories: the agglomerative methods in which we calculate outdistances between vertices and then incorporates the nearest points; and the divisive methods in which we take the whole graph and delete repeatedly arcs, dividing then the graph in new connex component.
\\ \\
Nevertheless, in some contexts of application of these methods of partition of graphs, the fact of proposing strict partitions, where a node of the graph is associated to a single cluster, might appear unsuited. Our research topics on the partition of graphs are applied in the field of the management, where the final objective consists in - roughly speaking - discovering "professional communities" starting from the analysis of graphs created from the logs of electronic exchanges between the employees. In this type of application, a desired property is to obtain at the end "overlapping partitions" from the graphs, i.e. partitions where a node - representing in our application an employee of the company - can be possibly associated to several communities. For instance, it is common to notice, in companies implementing matrix organisation structure, employees who take part in several projects and so, whom belong to several professional communities simultaneously.
\\ \\
There are only a few algorithms that basically have capacity to find out overlapping communities [09] [04]. Nevertheless, we have to notice that such kind of results can also be obtained thanks to the aggregation of partitions calculated with data arbitrary modified [02] or calculated by non-determinist algorithms.

\section{Our algorithm}
Proposed algorithm is neither divisive nor agglomerative. It belongs to a new kind of algorithms in two phases consisting in first extracting subsets of linked points from the graph and then incorporating those subsets in stable communities. We call it Pull \& Aggregate algorithm (P\&A algorithm).

\subsection{Extracting subsets of points...}
This first phase lists subsets of vertices broadly linked one to the others, subsets that will be used in the second phase to create communities. Different methods are possible to find those subsets of points out, for instance, methods through simulation of random walks. But another method was chosen, determinist and rather inexpensive in calculations: barycentric method.
\\ \\
In order to physically represent oneself the system, we have to imagine a set of pivots representing vertices linked to another by springs of null initial length and of stiffness defined by the value of the arc. The principle is to apply to this physical system a number of specific or uniform external forces. To calculate the positions of equilibrium, we use work of Huberman on the electric potentials [06]. It shows indeed that an iterative algorithm of successive calculation of weighted averages allows the resolution of the equations of Kirchhoff in a number of operations regardless of the number of vertices. We will use this type of algorithm for the calculation of the positions of the pivots in dimensions in which only specific forces are exerted, and a slightly modified algorithm of comparable nature when it is a question of also taking into account uniform forces.
\\ \\ 
Within large graphs, we work in two dimensions: a dimension according to X (specific forces), and a dimension according to Z (specific forces and uniform forces). For the smaller graphs, some adaptations are necessary - addition of a dimension Y, choice of several poles - but this is not the point of this present article.

\subsubsection{Equilibrium position according to X}
For dimension X, we successively consider each vertex of the graph. With a step i, we note this point Ai and fix its value to 1. We also fix at a value 0 all the vertices distant of more than n steps or diluted more than a x breakpoint in the case of graphs valued. To find these points quickly, we use BFS algorithm for not valued arcs and the algorithm of Dijkstra for valued arcs. All the other points are mobile. We then repeatedly calculate by weighted averages the equilibrium position of all the mobile points until reaching a steady equilibrium. Huberman shows that this phase of the algorithm is carried out in a time which does not depend on n. We can indeed make the analogy for calculations between the laws defining the value of electrical current in each node of an electronic circuit and the principle of the equilibrium of the forces: sum of forces is null and sum of intensities is null in each node; I = 1/R*U and the strength of a spring F = k*L; the concept of the voltage (difference of potential) is homogeneous at the distance (difference of length).

\subsubsection{Equilibrium position according to Z}
It is with this phase that the innovation compared to the basic barycentric methods is. It may be indeed in any cases that points not having anything to make with the cluster drawn towards the pole are found located at the same position [see figure \ref{f.2}]. If we base ourselves only on a computation of distances to determine the contents of the clusters, such points are thus added wrongly.

\begin{figure}[!h]
\begin{center}
\centerline{\epsfig{file=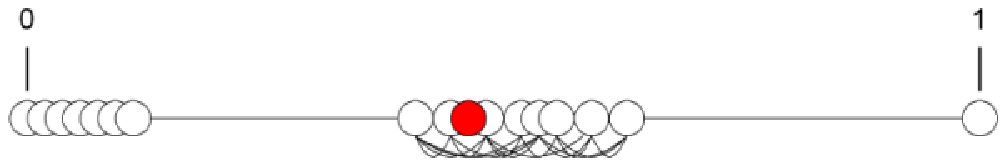,width=80mm}}
\caption{Diagram of a subset including a point located inopportunely at the same position}
\label{f.2}
\end{center}
\end{figure}

To locate these points, the idea is to apply an external force according to Z which indicates if each point is overall dependent with close points according to dimension X. The computation of the positions of equilibrium of systems including uniform external forces can't avoid the resolution of systems of equations, which implies an important complexity. We are thus constrained to escape from physical reality and propose a model which have the same advantages but which the solving is of a linear complexity. We preserve the same definition of fixed and mobiles points that in the computation of coordinates of X. The value of fixed points is 1, the one of the mobile points is initialized to 1 too. We repeatedly calculate the coordinates according to Z of each mobile point by weighted average by the value of the links and the distance of the neighbours. The difference between the position with the preceding iteration and the new position is noted d1. To represent the uniform force, we apply to the mobile point an additional displacement d2.
\\ \\ 
Additional displacement d2 is related to X and Z.
\\
d2 (X, Z) = d2x (X) * d2z (Z)
\\ \\ 
Several choices of functions are possible for d2x and d2z. d2x is choosen so that only the points having the greatest probability of belonging to the community of the pole fall under the effect of the external force and d2z so that the points tend to be stuck at the boundaries. The interval of displacement following Z is limited between 0 and 1: any final position not included in this interval is brought back to the nearest bound.
\\ \\
For pratical puposes, we retained for d2x a sigmoid function [see figure \ref{f.3}] and for d2z a square function [see figure \ref{f.4}].

\begin{figure}[h]
  \hfill
  \begin{minipage}[t]{.48\textwidth}
    \begin{center}  
      \epsfig{file=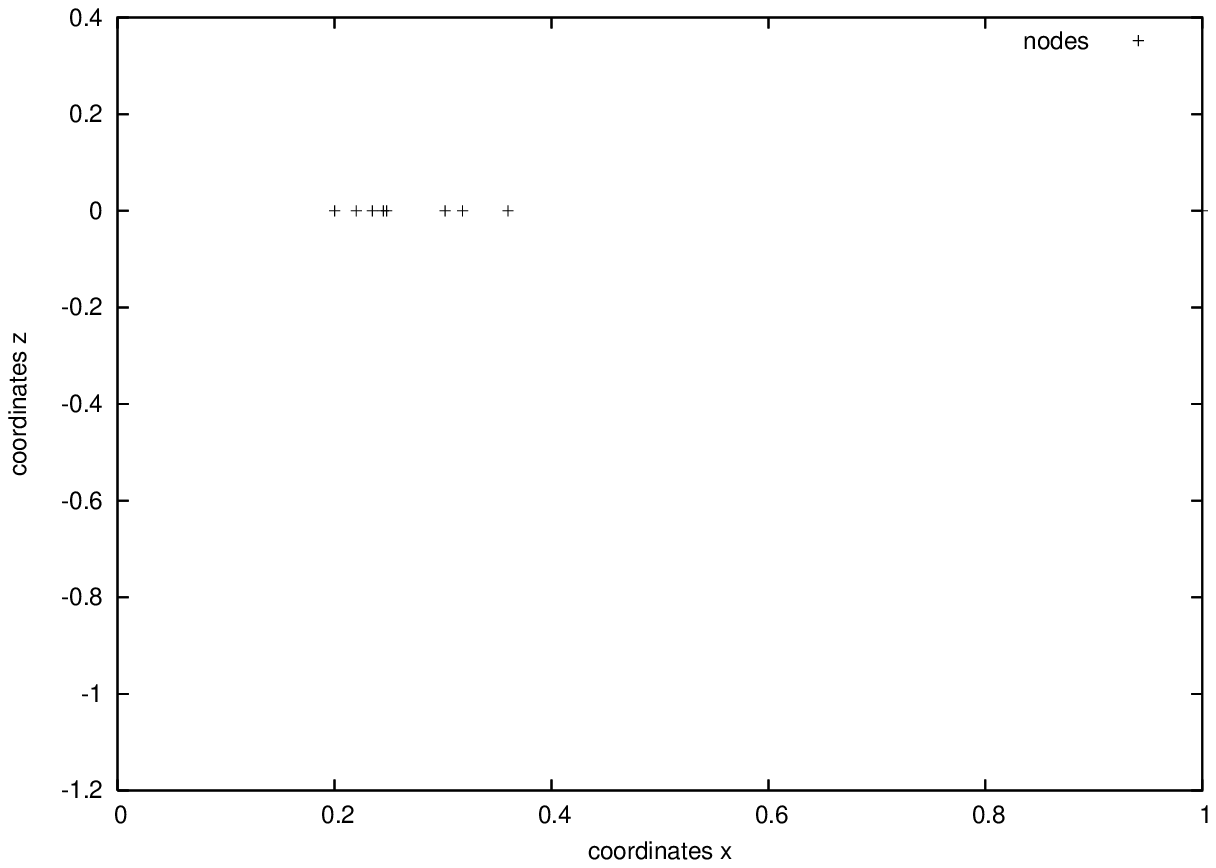,width=55mm}
      \caption{Sigmoid function}
      \label{f.3}
    \end{center}
  \end{minipage}
  \hfill
  \begin{minipage}[t]{.48\textwidth}
    \begin{center}  
      \epsfig{file=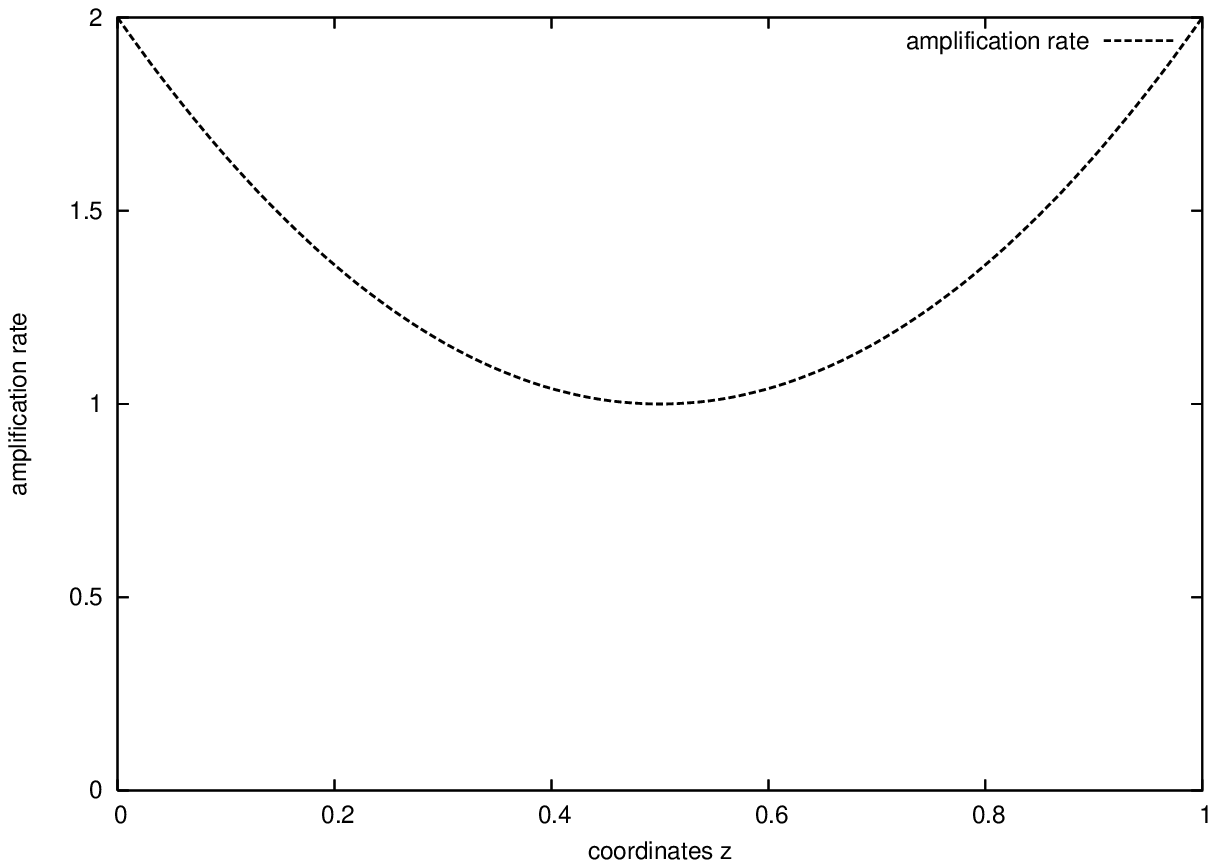,width=55mm}
      \caption{Square function}
      \label{f.4}
    \end{center}
  \end{minipage}
  \hfill
\end{figure}

The parameters of the sigmoid function depend on the number of points expected in our communities. If the estimated size of the communities is completely unknown or very variable, we can also test several values and retain that is before a jump in term of a number of points selected in the subset [see figure \ref{f.5}].

\begin{figure}[!h]
\begin{center}
\centerline{\epsfig{file=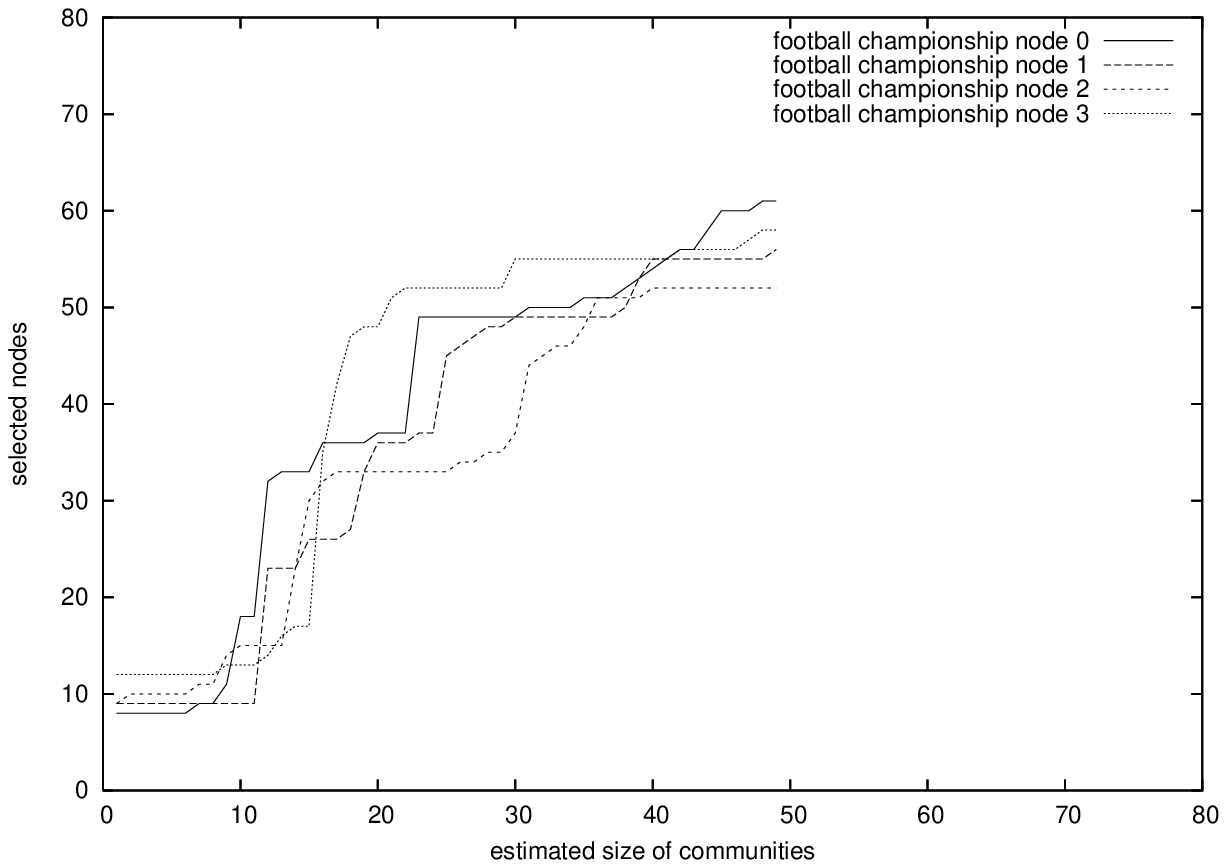,width=95mm}}
\caption{Diagram of the number of effectively selected points according to the estimated size of communities}
\label{f.5}
\end{center}
\end{figure}

In equilibrium, the subset of points to retain is consisted of all the points whose coordinate Z is lower than a threshold [see figure \ref{f.6}].

\begin{figure}[!h]
\begin{center}
\centerline{\epsfig{file=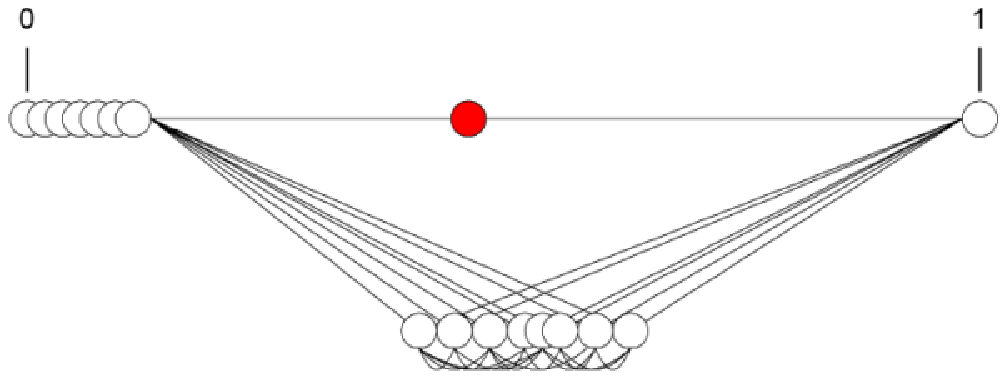,width=80mm}}
\caption{Diagram of the same subset including a point located inopportunely at the same position but this time with coordinates Z}
\label{f.6}
\end{center}
\end{figure}

Note: We do not have the theoretical proof that the algorithm using such functions converges and even less in a time not depending on n. Nevertheless, multiple tryouts indicate that coordinates according to Z converge in a few tens of iterations, whatever the size of the problem.

\subsection{Gathering the subsets of points in stable communities}
The second part of this algorithm is inspired of an algorithm of Huberman [05] modified to take into account the fact that we do not have as input a list of partitions of a graph, but only subsets of points forming subdivisions of communities.

\subsubsection{Determination of attractive poles for initialization}
It is a matter of defining, utilizing Huberman's words, a "basic masterList", some sort of skeleton of the main components of each community. We proceed by simple counting. Let us imagine that on the whole of the subsets, item X is present n time. It is maximum frequency of a point. We calculate for each other point Y, how many times Y is in the same subset that X. We put together the Yi points which are more than 50\% of the times together with X: We have just constituted first attractive pole. Among the remaining points, we choose the point presenting the most occurrences and we proceed in the same way to determine the second attractive pole. We continue until the list is empty or no more attractive pole found has a sufficient size (size significatively different than the size of the attractive poles previously found). Thus, the points remaining at the end, not associated to any attractive poles will be either points not very related to the communities or points related in an identical way identical to several communities. However that may be, they will be difficult to classify. 

\subsubsection{Fastening of the subsets of points to the attractive poles}
The continuation of the algorithm is almost identical to Huberman's one, with this difference that the formula retained for the measurement of proximity takes into account the non-homogeneity of the number of occurrences of the points: indeed, in the case treated by Huberman, the numbers of occurrences of each point are equal because a point is just represented once in each partition. For each group of points, we thus calculate its proximity with each community of the masterList then we amalgamate the group at its nearest community. This operation has a complexity of O(n), in fact we restrict this computation of proximity to the only communities that include at least one of the points of our group. The value of t is then the number of times where the element of the masterList was selected to combine. Once all the groups of points have merged with the attractive poles, we have to remove the points defined in an artificial way as attractive poles at the time of the initialization then calculate the relative part of presence of each point in each community.
\\ \\ 
Note: This phase of fastening to attractive poles may sometimes produce different results functions of the order of presentation of the subsets. This order being arbitrary, we have chosen to repeat the operation of aggregation a great number of times with variable orders of presentation then to make an average by attractive pole in order to obtain stable results (the list of attractive poles being constant regardless of the order of presentation).

\section{Applications}
We will quickly show the results obtained on simple graphs of average size (a few hundreds of points). These graphs are not valued and do not have, a priori, a structure of overlapping communities. We thus use in these two cases our algorithm to make partitioning. This one producing fuzzy communities, we consider that the partitions are made up of the points which belong to the communities with a maximum rate.

\subsection{Computer-generated graphs}
We test the performance of our P\&A algorithm on networks constructed with 192 vertices, divided into 12 separate communities of equal sizes, whom the average degree of vertices is 16. We make varying intra-community connections and determine how our algorithm performs in the cut of the graph in each case. To do so, we define a "success rate" representing the percentage of concord between the graph associated to the solution produced by the algorithm and the one produced by the parameters which allow us to build the computer generated graph.
\\ \\ 
As we can see on figure \ref{f.7}, the algorithm works perfectly when the number of intra-community connections is greater than the number of inter-community connections (intra 8.5 - inter 7.5 - success rate: 100\%). It's a fairly good result for the algorithm.

\begin{figure}[!h]
\begin{center}
\centerline{\epsfig{file=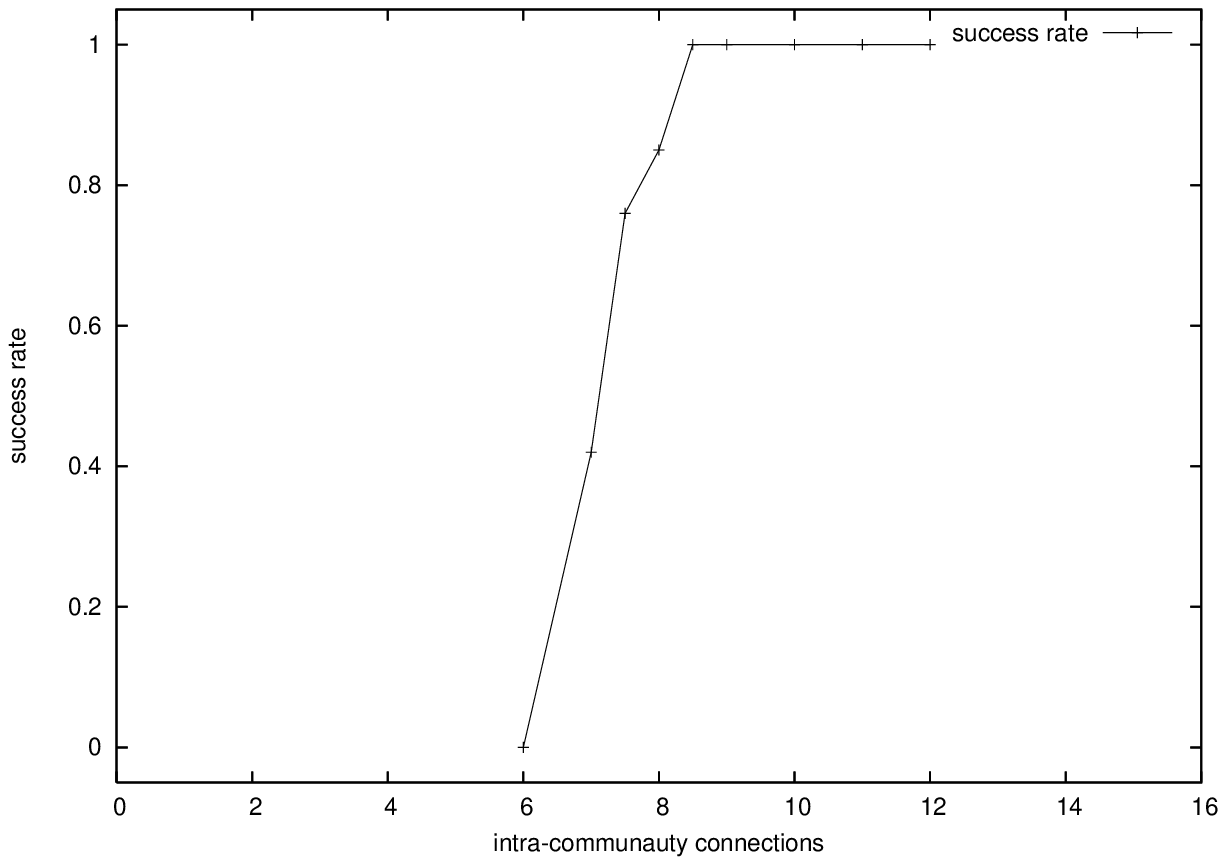,height=60mm}}
\caption{Success rate of the P\&A algorithm functions of intra-community connections of computer-generated graphs}
\label{f.7}
\end{center}
\end{figure}

\subsection{Football Championship}
We tested our algorithm on the data file "College football 2000" which was already analyzed amongst other things by Newman [01] and Radicchi [07]. We chose the spring as model of force and a neighbourhood of 2 steps for the determination of the mobile points.
\\ \\
See figure \ref{f.8} for results obtained by P\&A algorithm (noted PA), compared with of Newman's (noted GN) and Radicchi's (noted RA) and figure \ref{f.9} for complete results...
\\ \\
From this results, we can read for instance that node 84 belongs to community 2 at 87\% and to community 8 at 13\%. We can thus locate the nodes which belong to two or several communities that algorithms of strict partitioning would arbitrarily allot to one of them...  For example, according to the algorithm P\&A, node 50 belongs at 54\% to the community H and at 44\% to the community J. Twice algorithms GN and RA allot it to the community J, but they could have allotted it to H... We also notice that P\&A algorithm classifies all the points, contrary to the two other algorithms.

\begin{figure}[!ht]
\begin{center}
\centerline{\epsfig{file=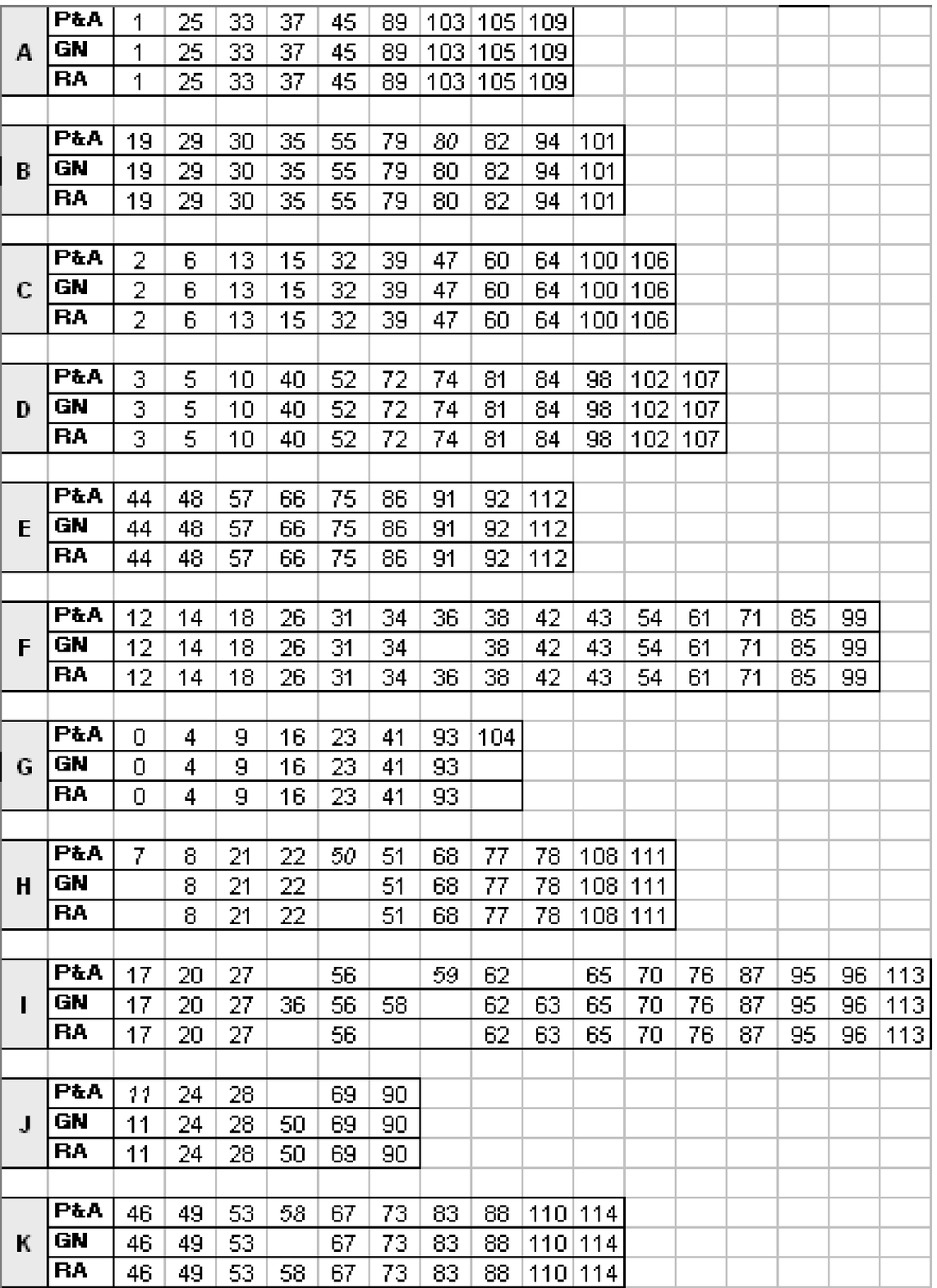,height=80mm}}
\caption{Comparison with GN's algorithm and Radicchi's}
\label{f.8}
\end{center}
\end{figure}

\begin{figure}[!ht]
\begin{center}
\centerline{\epsfig{file=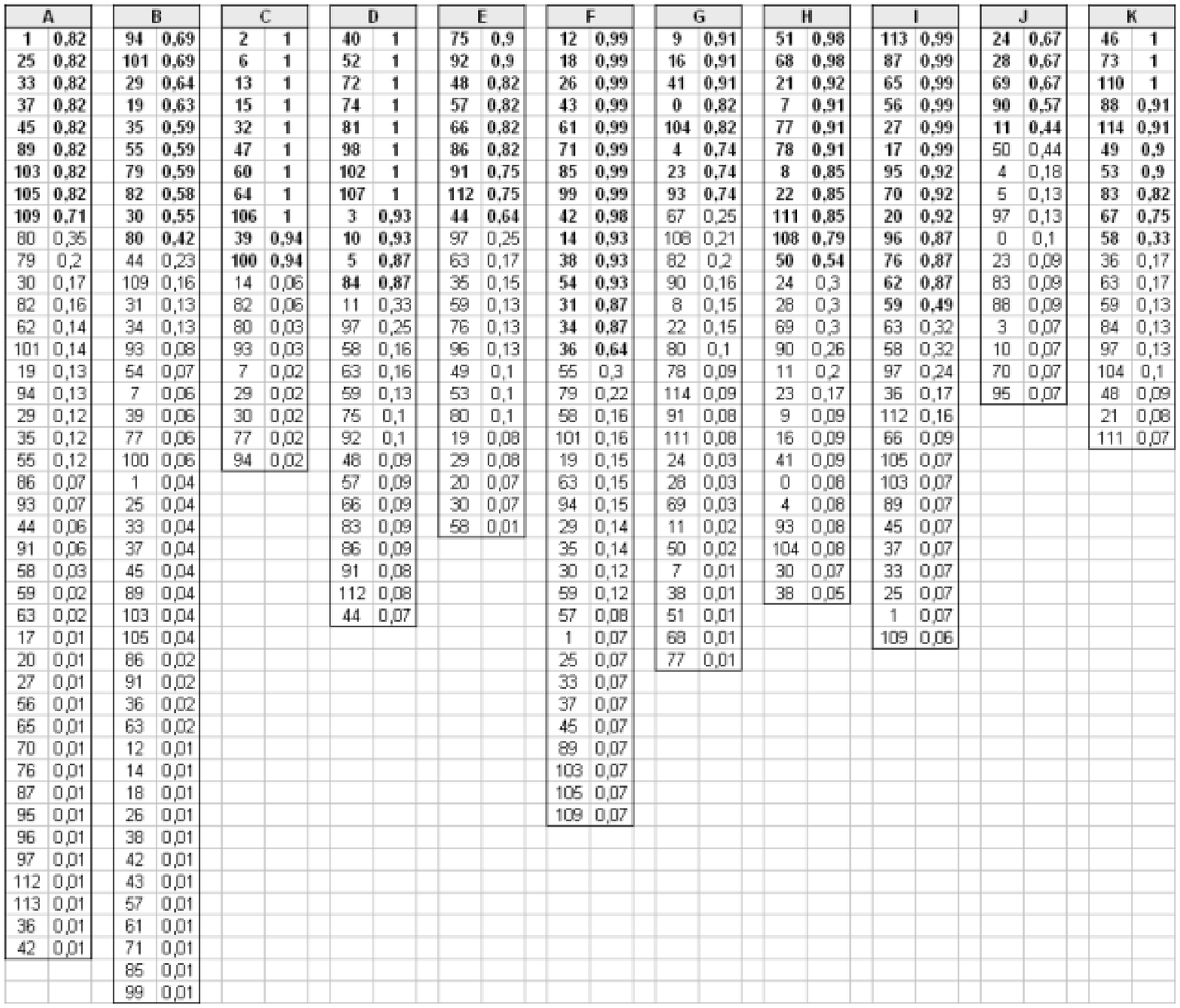,height=80mm}}
\caption{Complete results with percentage of belonging}
\label{f.9}
\end{center}
\end{figure}

\section{Conclusion and opening}
The P\&A algorithm, in addition to its low complexity which makes it perfectly adapted to calculations in very large graphs, presents a good number of other advantages.
\\ \\
First of all, the P\&A algorithm can be implemented and run in parallel, at least for the phase of extraction of the subsets which includes the greatest part of calculations: different machines or processors can compute their batch of pole in a completely independent way.
\\ \\
Moreover, the P\&A algorithm relying on local behaviour, it makes possible to detect the communities of delimited parts of the graph, without examining the wholeness of the graph. To do so, all we need to do is to select, as poles, points we want to associate to a community along with their n steps neighbours.
\\ \\
Lastly, the P\&A algorithm is more easily adaptable than purely mathematical methods: choice of thresholds, functions (specific forces: springs, rubber bands, models with rupture... | uniform forces: linear, sinusoid, sigmoid...) while preserving effective default values. It is thus particularly adapted to the networks with complex interactions (social networks for example). It should nevertheless be noted that according to the forms of forces, there is no formal proof that the algorithm converges and even less in a linear time. We can only see that on practical cases.
\\ \\
To go further, we will soon propose a more significant number of applications of this algorithm on real valued graphs  which include overlapping communities, such as social networks extrapolated from graphs of communications. We will apply other models of forces covering possible sociological significances. 

\section{Thanks}
We'd like to thanks Pascal Pons for providing us his program to generate random graphs and Mark Newman for the data on the football championship.

\section{Bibliography}

01 - M. Girvan and M. E. J. Newman (2001). Community structure in social and biological networks. Sante Fe Institute \& Cornell University.
\\ \\
02 - J. R. Tyler, D. M. Wilkinson, B. A. Huberman (2002). Email as Spectroscopy: Automated Discovery of Community Structure within Organizations. HP Labs.
\\ \\
03 - M. Girvan and M. E. J. Newman (2003). Finding and evaluating community structure in networks. Sante Fe Institute \& Cornell University.
\\ \\
04 - J. Baumes, M. Goldberg, M. Magdon-Ismail (2003). Efficient identification of Overlapping Communities. Rensselear Polytechnic Institute.
\\ \\
05 - B. A. Huberman and D. Wilkinson (2004). A method for finding communities of related genes. HP Labs.
\\ \\
06 - B. A. Huberman and F. Wu (2004). Finding communities in linear time: a physics approach. HP Labs.
\\ \\
07 - F. Radicchi, C. Castellano, et al. (2004). Defining and identifying communities in networks. The National Academy of Science of the USA.
\\ \\
08 - M. E. J. Newman (2004). Fast algorithm for detecting community structure in networks. University of Michigan.
\\ \\
09 - J. Reichardt and S. Bornholdt (2004). Detecting fuzzy community structure in Complex Networks with a Potts Model.
\\ \\
10 - J. P. Bagrow and E. M. Bollt (2004). A local method for Detecting Communities.
\\ \\
11 - L. Donetti and M. A. Muñoz (2004). Detecting network communities: a new systematic and efficient algorithm.
\\ \\
12 - M. Latapy and P. Pons (2005). Computing communities in large networks using random walks.
\\ \\
13 - http://en.wikipedia.org/wiki/Breadth-first\_search
\\ \\
14 - http://en.wikipedia.org/wiki/Dijkstra\'s\_algorithm

\end{document}